\documentclass{article}
\usepackage{graphicx}%
\usepackage{multirow}%
\usepackage{amsmath,amssymb,amsfonts}%
\usepackage{amsthm}%
\usepackage{mathrsfs}%
\usepackage[title]{appendix}%
\usepackage{xcolor}%
\usepackage{textcomp}%
\usepackage{manyfoot}%
\usepackage{booktabs}%
\usepackage{algorithm}%
\usepackage{algorithmicx}%
\usepackage{algpseudocode}%
\usepackage{listings}%
\usepackage[T1]{fontenc}
\usepackage{chemformula}
\usepackage{authblk}

\begin{document}

\title{
    \vspace{-1cm}
    Synthesis of epitaxial magnetic pyrochlore heterojunctions
}


\author[1, *]{Mikhail Kareev}%

\author[2]{Xiaoran Liu}

\affil[1]{Department of Physics and Astronomy, Rutgers University, Piscataway, New Jersey 08854, USA}

\affil[2]{Beijing National Laboratory for Condensed Matter Physics and Institute of Physics, Chinese Academy of Sciences, Beijing 100190 China}

\author[1]{Michael Terilli}

\author[1]{Fangdi Wen}

\author[1]{Tsung-Chi Wu}

\author[1]{Dorothy Doughty}

\author[3]{Hongze Li}

\author[3]{Jianshi Zhou}

\affil[3]{Department of Mechanical Engineering, University of Texas at Austin, Austin, Texas 78712, USA}

\author[2]{Qinghua Zhang}

\author[4]{Lin Gu}

\affil[4]{Beijing National Center for Electron Microscopy and Laboratory of Advanced Materials, Department of Materials Science and Engineering, Tsinghua University, Beijing 100084, China}

\affil[*]{Corresponding author: mikhail.kareev@rutgers.edu}

\author[1]{Jak Chakhalian}


\maketitle

\abstract{
The synthesis of stoichiometric and epitaxial pyrochlore iridate thin films presents significant challenges yet is critical for unlocking experimental access to novel topological and magnetic states.  
Towards this goal, we unveil an \textit{in-situ} two-stage growth mechanism that facilitates the synthesis of high-quality oriented pyrochlore iridate thin films.
The growth starts with the deposition of a pyrochlore titanate as an active iso-structural template, followed by the application of an \textit{in-situ} solid phase epitaxy technique in the second stage to accomplish the formation of single crystalline, large-area films.
This novel protocol ensures the preservation of stoichiometry and structural homogeneity, leading to a marked improvement in surface and interface qualities over previously reported methods. The success of this synthesis approach is attributed to the application of directional laser-heat annealing, which effectively reorganizes the continuous random network of ions into a crystalline structure, as evidenced by our comprehensive analysis of the growth kinetics. This new synthesis approach advances our understanding of pyrochlore iridate film fabrication and opens a new perspective for investigating their unique physical properties.}

\maketitle

\newpage

Rare-earth pyrochlores have emerged at the forefront of research due to their rich display of physical phenomena \cite{gardner2010magnetic}. Among these, spin ices pyrochlore titanates, A$_2$Ti$_2$O$_7$ (where A represents a lanthanide element), stand out. These materials are characterized by the lanthanide sites, which harbor strong frustration due to the constraints imposed on magnetic moments within the pyrochlore lattice. This frustration can result in a unique quantum spin-disordered state, marked by Coulomb correlations. Within this state, the concepts of emergent "magnetic monopole" and U(1) gauge excitations have garnered significant theoretical and experimental interest \cite{udagawa2021spin}. While classical spin ices such as Ho$_2$Ti$_2$O$_7$ and Dy$_2$Ti$_2$O$_7$ have been the center of attention, other rare-earth A$_2$Ti$_2$O$_7$ compounds, notable for their reduced effective spin and magnetic anisotropy, promise equally unconventional behaviors \cite{gingras2014quantum}.
A parallel line of interest lies with pyrochlore iridates, A$_2$Ir$_2$O$_7$, proposed to display topological phenomena driven by correlated electron dynamics including chiral spin liquids, axion insulators, and Weyl semimetals \cite{udagawa2021spin}. However, synthesizing large-scale single crystals of these compounds remains a challenge, sparking a keen interest in the epitaxial thin-film growth of pyrochlore iridates.

While the layer-by-layer growth of pyrochlore titanates has been recently well-documented \cite{wen2021epitaxial,wen2022correlated}, creating pyrochlores that include platinum group metals (such as Ru, Rh, Pd, Os, Ir, Pt) poses significant challenges. The \textit{in-situ} growth of pyrochlore iridate (PyIr) thin films, in particular, encounters obstacles due to the low reactivity of Ir at lower temperatures and the propensity for volatile species like IrO$_3$(g) to form from IrO$_2$(s) under high temperatures and O$_2$ pressures \cite{singh2022crystal}. To navigate these issues, several creative strategies have been employed, including physical vapor deposition (PVD) with co-sputtering of IrO$_2$ \cite{guo2021searching}, solid phase epitaxy (SPE thereafter) \cite{kim2022perspective}, and flash annealing methods \cite{kim2020strain,kim2019operando,song2023engineering}. Notably, the recently-established SPE technique, devoid of any additional growth components, offers a simplified route for \textit{in-situ}  growth of epitaxial pyrochlore thin films and heterostructures \cite{liu2020situ}. Furthermore, the epitaxial growth by pulsed laser deposition (PLD) enables precise control over the valence state of the cation sublattice by finely tuning the oxygen partial pressures within a narrow growth window.\cite{wen2021epitaxial,wen2022correlated}. Such tunability is important, as in the SPE process, the successful formation of the pyrochlore phase hinges on precise oxygen (O$_2$) control during the initial amorphous layer deposition and the subsequent post-annealing crystallization. 


Despite these advancements in fabrication techniques, SPE-grown pyrochlore films exhibit surface corrugation. As a result, films grown by this method are largely incompatible with the application of surface-sensitive analytical methods, such as Vacuum Ultra-Violet Angle-Resolved Photoelectron Spectroscopy (VUV-ARPES) and Scanning Tunneling Microscopy (STM), crucial for probing their topological and electronic properties. This limitation underscores the need for ongoing efforts to refine growth protocols, aiming to markedly improve the surface and interface quality of pyrochlore films.

\begin{figure}[ht]%
\centering
\includegraphics[width=0.9\textwidth]{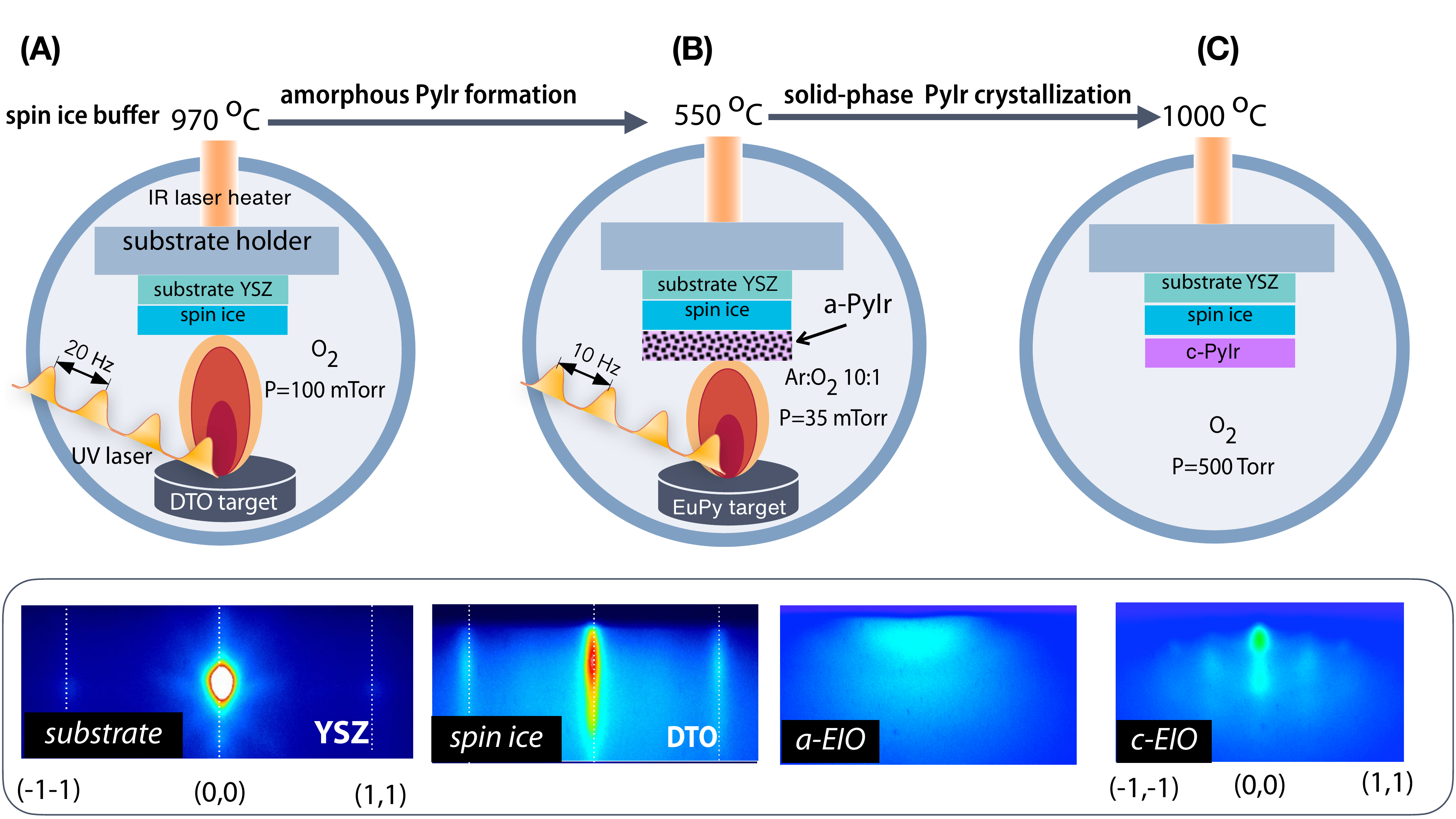}
\caption{Pyrochlore iridate film fabrication. (a) DTO grown layer-by-layer BY PLD on yttria-stabilized zirconia (YSZ), (b) deposition of amorphous EIO pyrochlore layer, then (c) SPE at a high-temperature annealing synthesis cycle}\label{fig1}
\end{figure}


In this study, we introduce a new approach for synthesizing high-quality pyrochlore iridate thin films, marking an advancement in the sub-field. Leveraging the strengths of our previously-developed SPE method \cite{liu2020situ,liu2021magnetic}, we incorporate a pyrochlore titanate, Dy$_2$Ti$_2$O$_7$ (DTO), as an active iso-structural template \cite{wen2021epitaxial,wen2022correlated}, combined with the directional laser-heat annealing, that effectively reorganizes the continuous random network of ions into a crystalline, stoichiometrically-perfect pyrochlore structure. The new protocol not only enhances the interfacial quality, as confirmed by Scanning Transmission Electron Microscopy (STEM) imaging and High-Angle Annular Dark Field (HAADF)/Electron Energy Loss Spectroscopy (EELS) elemental analysis, but also remarkably improves the surface quality, as verified by X-ray reflectivity and Atomic Force Microscopy (AFM) results. This materials discovery opens exciting opportunities for in-depth experimental exploration of the topological, electronic, and magnetic characteristics inherent to pyrochlore iridates and spin ice compounds, offering promising venues for research with artificial quantum architectures.

\section{Experiment and Results}\label{sec2}

In what follows, we present a comprehensive description of our new experimental design for the SPE procedure. Specifically, we have replaced the traditionally-employed yttria-stabilized zirconia (\ch{Y2O3:ZrO2}, YSZ) substrate with a DTO pyrochlore active \textit{iso-structural} template. This substitution explicitly aims to investigate the role of the anion (oxygen) sublattice in synthesizing pyrochlore iridates. The DTO template is deposited on YSZ substrates in a layer-by-layer fashion described in references \cite{wen2021epitaxial,wen2022correlated}. Following the deposition of DTO, a layer of Eu$_2$Ir$_2$O$_7$ (EIO) is synthesized atop the DTO block via a two-step process: (i) deposition of amorphous EIO pyrochlore layers, followed by (ii) SPE facilitated \textit{in-situ} through a high-temperature annealing cycle \cite{liu2021magnetic}. In this context, SPE is conducted on the pyrochlore DTO template rather than the commonly used fluorite (YSZ) structure. Our dual-stage annealing protocol, comprising PLD-SPE, leverages the formation of IrO$_3$ volatiles at elevated oxygen pressures from solid IrO$_2$. The PLD part of the synthesis is meticulously calibrated to achieve the desired film thickness by precisely adjusting laser pulse counts. As a result, at the first stage, an amorphous pyrochlore matrix free from iridium deficiencies is formed \cite{liu2020situ,liu2021magnetic}.

Subsequently, the film undergoes post-annealing, with growth parameters detailed in Fig. \ref{fig1}. Post-annealing duration is a critical tuning parameter: short post-annealing intervals result in a mixture of solid and amorphous phases, whereas extended post-annealing intervals lead to phase separation. As seen in Fig. \ref{fig1}b during the deposition phase, reflection high-energy electron diffraction (RHEED) imaging reveals broad rings, indicative of short-range order established within the amorphous pyrochlore layer, as opposed to diffusive scattering patterns characteristic for a completely disordered RHEED halo \cite{jeong2003ring}. Although RHEED monitoring is infeasible during our high-pressure SPE annealing process (about 500 Torr of oxygen), the post-annealing RHEED images exhibit a characteristic streaky pattern, implying the formation of long-range crystalline order [see Fig. \ref{fig1}c].

\begin{figure}[ht]%
\centering
\includegraphics[width=0.9\textwidth]{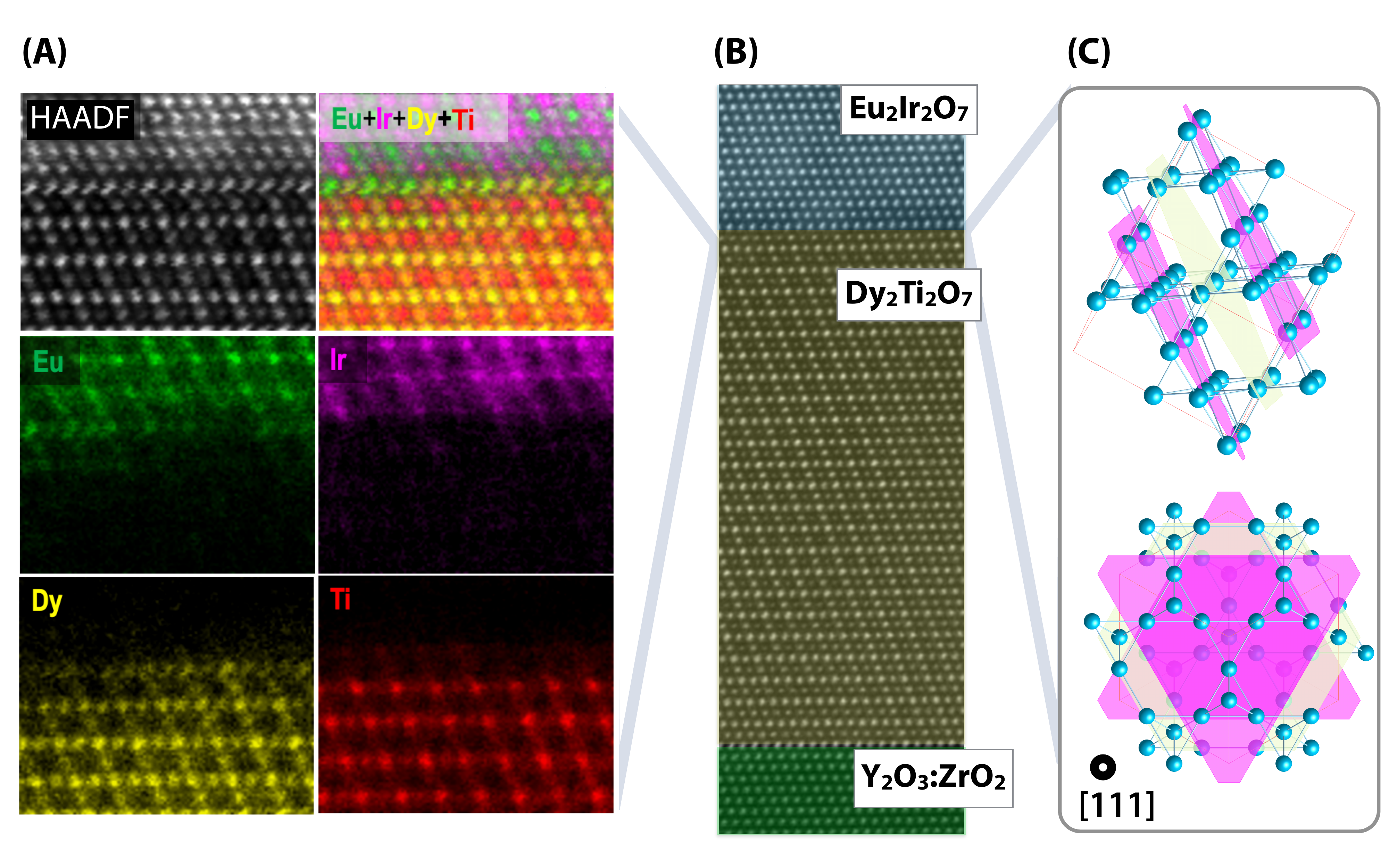}
\caption{ Cation (A-site and B-site) sublattices, Kagome-triangular sequence and pyrochlore phase can be visualized by; (a) elemental mapping based on STEM-EELS analysis, (b) STEM-HAADF, (c). Kagome-triangular sequence}\label{fig2}
\end{figure}

\begin{figure}[ht]%
\centering
\includegraphics[width=0.9\textwidth]{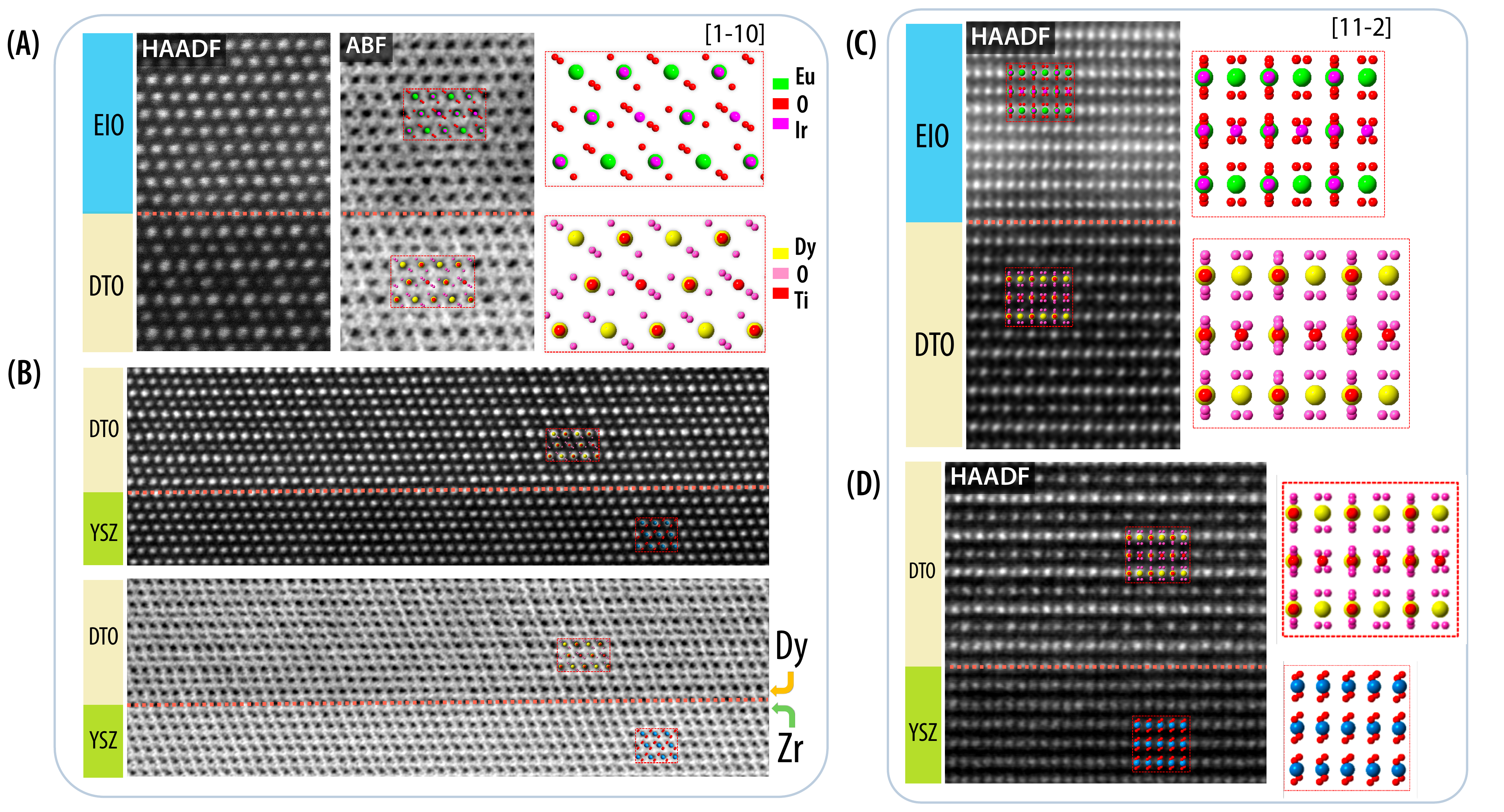}
\caption{HAADF and ABF image comparison  to identify the oxygen sublattice along the [1-10] and [11-2] projections ( (a-b and c-d accordingly). The stacking sequence at the Py-Py (EIO-DTO) and Py-Fl (DTO-YSZ) interfaces identified in HAADF-ABF images along the [11-2] and [1-10] projections.}\label{fig3}
\end{figure}

The formation of the pyrochlore amorphous matrix occurs during the initial PLD stage under low-temperature and low-pressure conditions in the Ar-O$_2$ gas mixture as a part of the pyrochlore iridate film's fabrication sequence [Fig. \ref{fig1}a]. The key question if the amorphous phase maintains pyrochlore stoichiometry is validated through several advanced analytical techniques \cite{liu2020situ,gallagher2016epitaxial,fujita2015odd}. To elucidate the films' crystallinity and electronic and chemical properties after post-annealing, we conducted detailed STEM-EELS and HAADF-ABF measurements and analyses. Post-annealing examinations shown in Fig. \ref{fig2} clearly demonstrate the absence of the amorphous phase, as evident in atomically resolved HAADF and ADF images. The pyrochlore phase and the A-site (Dy, Eu) and B-site (Ti, Ir) cation sublattices and Kagome-triangular sequence are visualized by elemental mapping based on the STEM-EELS analysis. Additionally, anion (oxygen) sublattice motifs are distinctly visible in the ABF images; atomic structure models are superimposed on these ABF images to identify the oxygen sub-lattice across the [1-10] and [11-2] projections (Fig.s \ref{fig2} a-b and c-d respectively). In addition, as shown in (Fig. \ref{fig3}c), the stacking sequence at both the EIO-DTO and DTO-YSZ interfaces is readily identified in HAADF-ABF images along the [11-2] and [1-10] projections, corroborated by the ABF analysis in reference \cite{mostaed2017atomic}.

A direct comparison of HAADF vs. ABF imaging reveals that the oxygen sublattices at the EIO/DTO interface are continuous and identical. 
Because of the specific design of our planar laser heat source, the DTO layer acts as the iso-structural template\cite{ohno2023template}, facilitating the SPE of amorphous EIO layer crystallization wave propagating toward the surface (\textit{directional} heat wave propagation). Assuming a conventional solid-state reaction (SSR) mechanism, one would expect to observe intermixing due to the diffusion of the Eu${2}$O${3}$ and IrO$_{2}$ components during the pyrochlore-forming SSR process. This effect is common for the interfacial reactions involving moving heterophase boundaries \cite{hesse2004ssr}. However, as indicated, the intermixing at the EIO-DTO interface is strictly confined to a few atomic planes, a statement supported by elemental mapping illustrated in Fig. \ref{fig2}a. As for the extent of chemical ion intermixing at the \ch{Dy2Ti2O7}/YSZ interface, particularly within the cation sublattice, it remains ambiguous due to the L-edge of Zr being beyond the detection energy range of the EELS spectrometer. On the other hand, the ABF imaging of the DTO/YSZ interface, as shown in Fig. \ref{fig2}, strongly suggests that the oxygen sublattices are in a close match between DTO (pyrochlore) and YSZ (fluorite).

Next, we turn our attention to the surface structural properties. For this purpose, we evaluated the surface roughness of both single-layered EIO and YIO and bi-layered EIO/DTO and YIO/DTO configurations, presented in Supplemental Table 1. Surface roughness was quantified using low-angle X-ray reflectivity (XRR) measurements, which were modeled to deduce the roughness parameters. First, we note that compared to single-layer films, the bilayer growth on DTO markedly improved the surface roughness by 34\% and 42\%, respectively. In addition, the critical-angle intensity is significantly higher for the bilayer films, indicating a flatter surface.
Furthermore, atomic force microscopy (AFM) scans were taken on both sets of films in a 1.5x1.5µm field of view to obtain a two-dimensional real-space measurement of the surface roughness. These measurements reveal an even more drastic difference in the surface roughness; we report an Sq value five times lower in bilayer EIO/DTO and nine times lower for YIO/DTO when compared to their single-layer counterparts. Combined, the XRR and AFM measurements confirm a remarkably lower roughness achieved when using our new DTO-templated PLD-SPE method.  

\section{Discussion}\label{sec3}

The selection of an appropriate substrate or buffer layer is critical for the effectiveness of our new annealing process. Commonly, solid-phase epitaxy (SPE) post-annealing is executed \textit{ex-situ in a furnace}, characterized by an extremely non-uniform propagation of heat. In stark contrast, our in-situ annealing approach places the substrate's back side in direct contact with a planar heat absorber powered by a 140 W fiber-coupled diode laser \cite{Surfacewebsite}. This specifically-designed configuration facilitates directional phonon propagation from the substrate to the interface, emphasizing the critical nature of the interface choice for our crystallization technique.

One significant reason for the DTO template layer's contribution to the markedly improved surface and interface quality is attributed to the formation of a pyrochlore-pyrochlore interface between titanate and iridate layers. Unlike the fluorite structure of YSZ substrate, which randomly hosts vacancies at one-eighth of its anion sites, the DTO/EIO bilayer ensures complete anion sublattice matching and fosters immediate nucleation at a submicroscopic level. Additionally, the atomic precision achieved in the layer-by-layer growth of DTO yields an atomically-flat interface, which promotes a significantly smoother iridate layer crystallization in the vicinity of the interface. The flatness of the interface propagates throughout the iridate layer during the SPE crystallization process, resulting in a low-roughness surface. This observation aligns with the previous report highlighting the role of the DTO-templated (111)-oriented YSZ in mitigating both lattice mismatch and structural dissimilarities \cite{sasaki2004atomic}.

Furthermore, we can conjecture that the DTO layer's inclusion significantly contributes to the success of the amorphous growth method. A recent thermal conductivity study on GeS/PbTe superlattices showed a substantially-reduced interfacial thermal resistance (ITR) when the GeS layer was amorphous, attributed to a large overlap in the vibrational mode density of states (DOS) between amorphous-GeS and crystalline-PbTe \cite{ishibe2021heat}. Analogously, the large phonon DOS overlap between DTO and crystalline YIO suggests a minimal interfacial thermal resistance, facilitating remarkably efficient heat diffusion across the interface during annealing. The process of  thermal diffusion is crucial for the rapid recrystallization of the amorphous layer into a pyrochlore structure. It is worth noting that the  disordered nature of doped YSZ makes phonon propagation through the DTO layer more facile compared to YSZ \cite{che2022thermal,schelling2001mechanism}.
Further, recent experiments using reflection high-energy electron diffraction to investigate the SPE kinetics have shown that crystallization occurs via the thermally-activated propagation of the crystalline/amorphous interface \cite{marks2021solid}. Given YSZ's known properties as a thermal barrier coating material with exceptionally low thermal conductivity, its influence on the SPE synthesis is deemed negligible here. Both YSZ and DTO exhibit similar thermal conductivity values, ensuring comparable heat wave distribution \cite{ren2015high,zhou2022thermal,tachibana2022thermal}.

It is natural to assume that the DTO template also plays a pivotal role in modulating the strain state of the pyrochlore iridate layer. Reciprocal space mapping indicates that the EIO layer exhibits partial strain (see Supplemental Fig. 6). As expected for the layer-by-layer growth, the DTO layer is fully strained by the YSZ substrate. Interestingly, unlike DTO, the iridate layer is only partially strained, underscoring the DTO layer's superiority over YSZ as an iso-structural template for iridate film growth. This observation of the partially-strained EIO templated by DTO opens a promising avenue for future work aimed at achieving comprehensive strain control. The presence of partially strained thin films provides a utility to manipulate lattice strain in iridates films, inducing potentially new topological states predicted by theory\cite{kareev2011sub}.

\begin{figure}[ht]%
\centering
\includegraphics[width=0.9\textwidth]{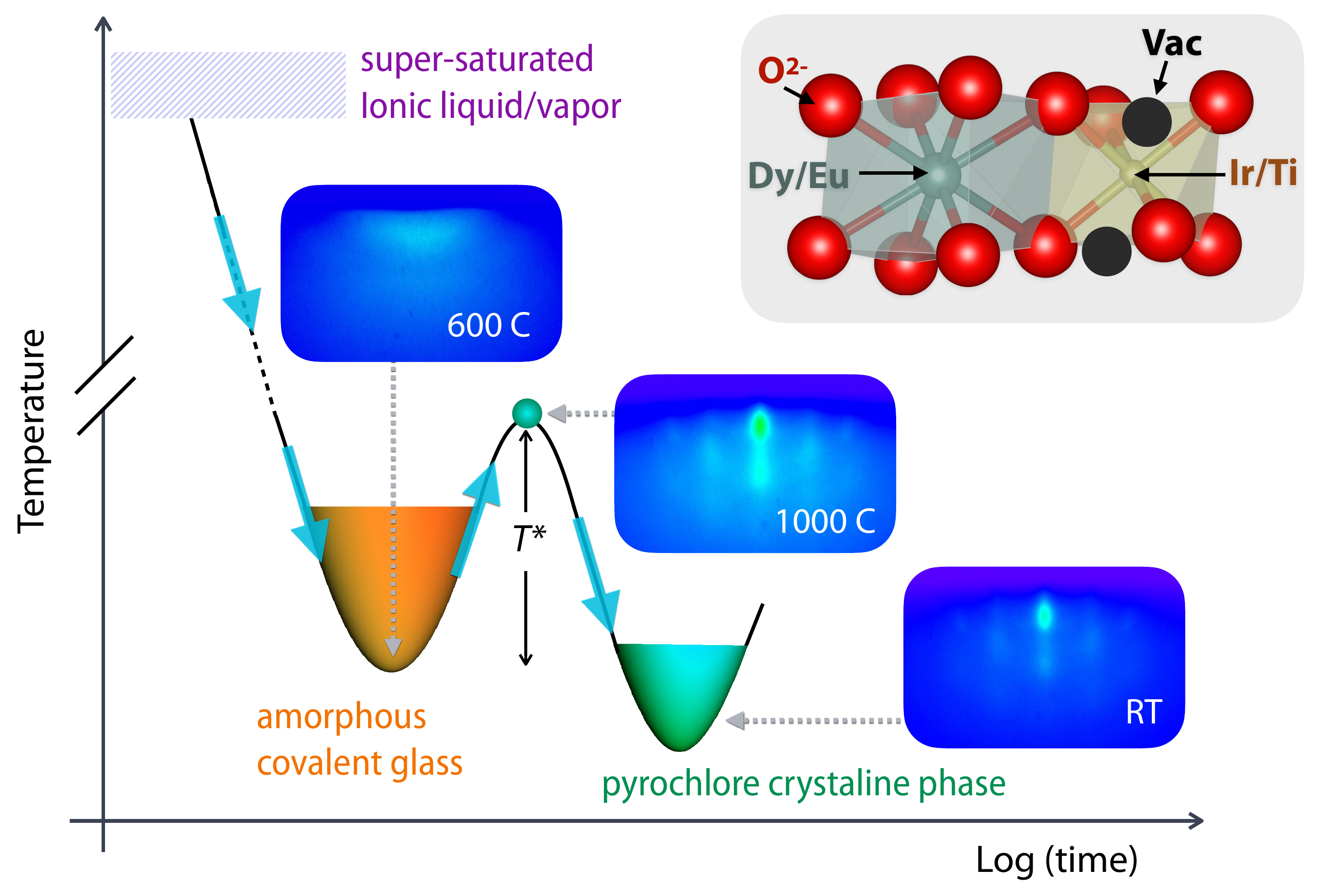}
\caption{Kinetic handle for the fine-tuned manipulation of growth regimes. Amorphous film ultrafast quenching at the pre-SPE stage (crystallization was suppressed, a metastable supercooled liquid (glass state) was formed. Devitrifying this random continuous network leads to the formation of pyrochlore iridate film (SPE crystallization) with excellent crystallinity.}\label{fig4}
\end{figure}

In what follows, we speculate about the kinetics of the new growth mode by PLD-SPE. SPE itself involves the ordering of atoms from a metastable amorphous phase into a crystalline structure via local bond rearrangements at the moving crystalline-amorphous interface upon heating, all occurring in the solid state \cite{johnson2015solid}.  At the start of the SPE crystallization, it enforces the amorphous material to adopt the lattice structure of the underlying crystalline substrate, serving as a template for growth \cite{johnson2015solid}. Next, we attempt to bring light to the correlation between the activation energy of SPE and the characteristics of the crystalline-amorphous interface that facilitates the bond rearrangement process.
Our two-stage fabrication protocol, depicted in Fig. \ref{fig4}, begins with the deposition of an amorphous stoichiometric matrix, followed by high-temperature post-annealing during the second SPE stage. Extensive RHEED studies on perovskite oxides grown by SPE  suggest that crystallization progresses via the thermally activated movement of the crystalline/amorphous interface  \cite{marks2021solid}. These results also suggest a consistent mechanism underlying PyIr thin film growth. Initially, PyIr films represent a kinetically stabilized, thermodynamically non-equilibrium amorphous solid or covalent glass, wherein the molecular disorder and the thermodynamic properties corresponding to the state of the respective under-cooled melt are frozen. This overall transformation process is illustrated in Fig. \ref{fig4}. During the film deposition, ultrafast cooling (quenching) prevents nucleation, maintaining the metastable film's amorphous state. Subsequent devitrification of this random continuous network yields a pyrochlore iridate film with superior crystallinity \cite{sambri2012plasma}.

As for the substrate, it plays the role of the `wetting' layer, which is crucial for SPE. In particular, the `thermodynamic interface penalty' model suggests that materials with inherent `glass forming ability' may preferentially crystallize at the crystal substrate surface, even as the bulk of the film remains amorphous \cite{russo2018glass}. This phenomenon is contingent upon the mobility of structural units, with recrystallization occurring at the buried crystalline-amorphous interface. The dynamics of such 2D phase transitions arising from dimensional confinement during SPE growth have been extensively studied in the past  \cite{gutzow1995vitreous, gutzow2011glasses, li2021phase}. Recent theory emphasized the role of  `bond-orientation ordering' hidden behind crystallization, underscoring its significance in a supercooled state that affects the kinetic pathway of crystallization process. \cite{tanaka2011roles, tanaka2019revealing}

In what follows, we delve into an intriguing question of the bond rearrangement phenomena occurring within pyrochlore lattices, as depicted in Fig. \ref{fig4} inset. The structure of these lattices is represented by a cubic cell centered around a cation, with oxygen atoms or vacancies positioned at the vertices. The connectivity among various cation sites is facilitated through the shared edges of these cubic cells. Notably, in pyrochlore structures, the cations Dy/Eu ($3+$) and Ir/Ti ($4+$) exhibit eight and six coordination numbers, respectively. This distinct arrangement leads to the emergence of a superlattice, defined by its orderly distribution of cations and vacancies. Such an arrangement promotes the development of polyhedra that share corners and edges, thereby defining a cubic unit cell designated as EIO or DTO (space group $Fd\bar{3}m$).
During the deposition process, integrating a new structural unit into the pre-existing random network is dictated by the connectivity of the existing units within the network. This geometrical strategy, known as topological modeling, aids in generating disordered networks, permitting variations in valence bond angles away from their crystalline equilibrium states. For glasses composed of anisotropic constituents, the energy cost associated with rotational movements renders them less favorable, which, in turn, restricts their translational motion. As a result, the vibrational properties of such a covalent network, including the phonon density of states, are expected to demonstrate variability in accordance with the degree of anisotropy of the constituent ions.
Within this framework, it is observed that low-energy rotational degrees of freedom, which operate independently of translational modes, induce a synchronous rotational movement among neighboring ions, yet without displacing them. This phenomenon bears significance for understanding thermal conductivity within the disordered (amorphous) state of pyrochlore iridates \cite{yunker2011rotational}.

Finally, we address the conditions inhibiting unwanted crystallization during the interrupted pulse laser deposition (PLD) process \cite{wen2021epitaxial,wen2022correlated,liu2020situ,kareev2011sub}. Adopting an interrupted PLD mode provides precise kinetic control, effectively postponing the onset of crystallization under conditions of extreme supersaturation and fast flux modulation \cite{kareev2011sub}. 
The formation of an amorphous film is facilitated by ultrafast quenching during the liquid phase of solid-phase epitaxy, thus averting crystallization and fostering a kinetically stabilized supercooled liquid state — a distinct manifestation of the glassy state \cite{cavagna2009supercooled}.
The formation of an amorphous film is achieved through ultrafast quenching at the liquid phase of the SPE, thereby suppressing crystallization and inducing a kinetically stabilized supercooled liquid state — essentially another manifestation of the glassy state \cite{cavagna2009supercooled}. Intriguingly, the kinetic behaviors observed under SPE conditions mirror those of the pre-SPE stage. In our procedure, a nonlinear cooling regime is set via rapid cooling at higher temperatures until the minimum nucleation threshold is reached; this is followed by a slower cooling rate at lower temperatures to accommodate the increasing relaxation time of the supercooled liquid. Conversely, maintaining a constant cooling rate results in premature crystallization at the interface with the template.\cite{ishibe2021heat,bolmatov2015unified,trachenko2023theory}.


In conclusion, we unveil an \textit{in-situ} two-stage growth mechanism that facilitates the synthesis of high-quality oriented pyrochlore iridate thin films. The first stage relies on the application of interrupted PLD mode that effectively postpones the onset of crystallization under extreme supersaturation and fast flux modulation and promotes the formation of the random continuous network followed by the direction laser annealing protocol. This novel method ensures the preservation of stoichiometry and structural homogeneity, leading to a remarkable improvement in surface and interface qualities over previously reported methods.
A detailed analysis of the growth kinetics supports the effectiveness of this approach, demonstrating how the deposition and annealing processes contribute to the successful synthesis of the films and heterojunctions of pyrochlores iridates.
This new advanced synthesis method not only enhances the quality of pyrochlore iridate thin films but also deepens our understanding of their synthesis. Furthermore, by improving the structural and electronic properties of these films, this growth method finally opens up new opportunities for exploring the unique physical properties of pyrochlore iridates, with implications for the discovery of new quantum topologically non-trivial states.



%
%


\newpage
\noindent{\bf Methods}\\
The PyIr targets were ablated using a KrF excimer laser (lambda = 248 nm, energy density about 3 J/cm2) at a repetition rate of 10 Hz. Unlike previously-reported work, our deposition was carried out at a much lower substrate temperature of 550 $^\circ$C, under a 35 mtorr atmosphere of a mixture of Ar and \ch{O2} gases (partial pressure ratio, Ar: \ch{O2} = 10:1). The film was post-annealed inside the chamber at 1000C, under a 500 Torr atmosphere of pure \ch{O2} (30 min for Eu, < 20 min for Y), and then cooled down to the room temperature. Detailed description of spin ice growth can be found elsewhere. \cite{wen2021epitaxial,wen2022correlated}. We used an X-ray diffractometer (Malvern Panalytical Empyrean) to characterize the reciprocal space of the sample and to understand the film's crystal structure to obtain (1) X-ray reflectivity (XRR) and (2) X-ray diffraction (XRD, reciprocal space map (RSM)) data. Apart from X-ray-related measurements, which are an important statistical view of sample quality, we have obtained the annular bright field (ABF) and high-angle annular dark field (HAADF) STEM-EELS on our films. This result was measured in collaboration with Lin Gu's group in the Institute of Physics, Chinese Academy of Science.


\noindent{\bf Supplementary Information}\\
This article contains Supplementary information.

\noindent{\bf Acknowledgements}\\
M.K., M.T., F.W., T.W., D.D., and J.C. acknowledge the support by the U.S. Department of Energy, Office of
Science, Office of Basic Energy Sciences under award number DE-SC0022160. X.L. acknowledges the support by the National Natural Science Foundation of China (Grant No. 12204521) and the National Key R$\&$D Program of China (Grant No. 2022YFA1403400). H.L. and J.Z. acknowledge the support from NSF Grant No. DMR-1720595 and DMR-2308817. Q.Z. and L.G. acknowledge the support by the National Natural Science Foundation of China (52250402,52025025).





\bigskip







\newpage
\bibliographystyle{naturemag} 
\bibliography{PyIr_growth} 
\newpage

\section{Supplementary Information}\label{secS1}

\setcounter{figure}{0}
\makeatletter 
\renewcommand{\thefigure}{S\@arabic\c@figure}
\makeatother

\begin{figure}[ht]%
\centering
\includegraphics[width=0.9\textwidth]{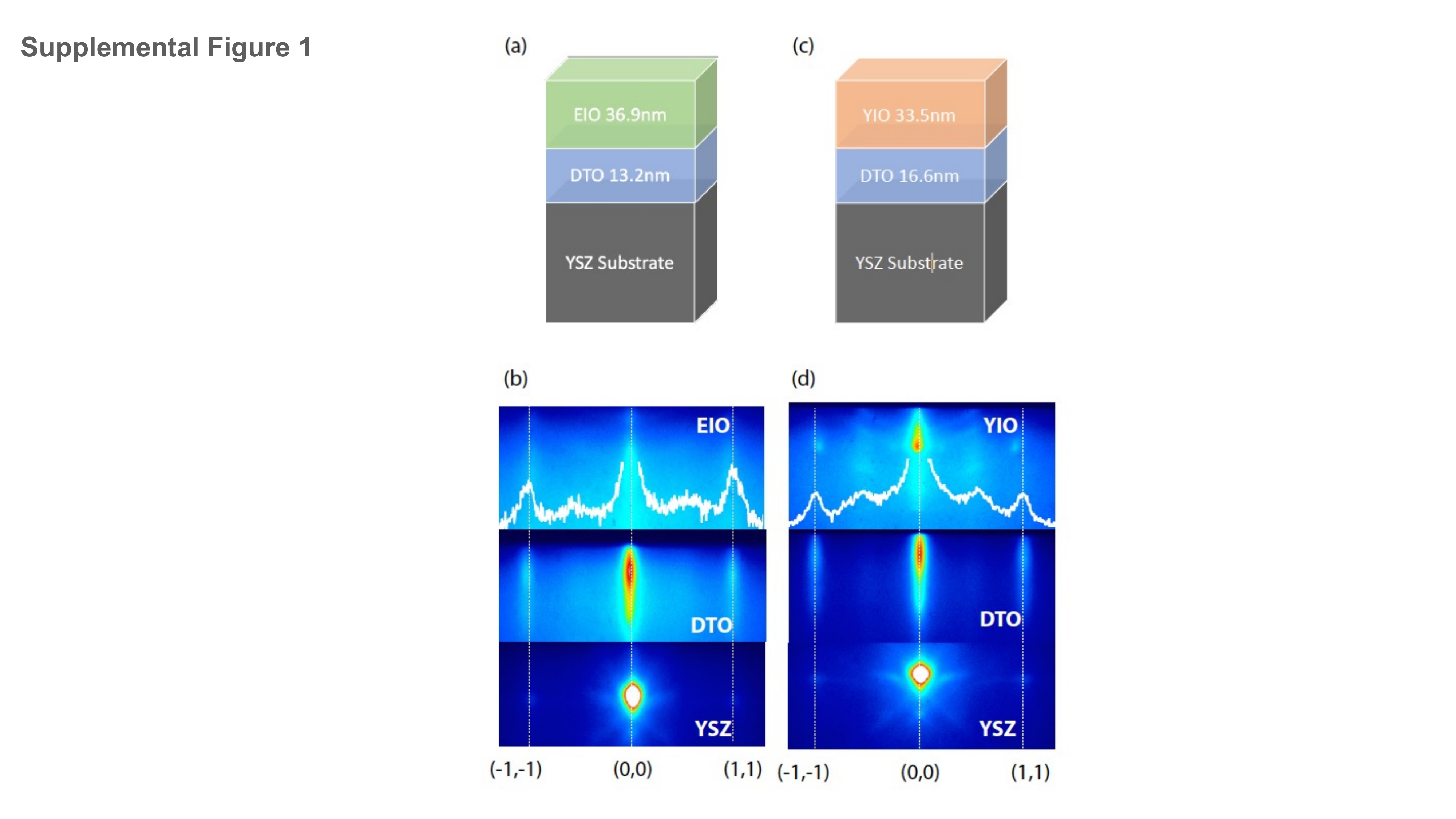}
\caption{(a) Schematic picture of \ch{Eu2Ir2O7}/\ch{Dy2Ti2O7} bilayer on the YSZ substrate. The thickness was obtained through XRR fitting. (b) RHEED images of substrate before deposition (bottom), after the deposition of \ch{Dy2Ti2O7} (middle), and after the deposition of \ch{Eu2Ir2O7}(top). (c) Schematic picture of \ch{Y2Ir2O7}/\ch{Dy2Ti2O7} bilayer on the YSZ substrate. The thickness was obtained through XRR fitting. (d)RHEED image of substrate before deposition (bottom), after the deposition of \ch{Dy2Ti2O7} (middle), and after the deposition of \ch{Y2Ir2O7} (top).}\label{S1}
\end{figure} 

\newpage

\renewcommand{\thetable}{S\arabic{table}}

\begin{table}
    \centering
    \begin{tabular}{cccc}
         Sample&  Thickness (nm)&  XRR roughness (nm)& AFM Sq (nm)\\
         \ch{Eu2Ir2O7}&  45(5)&  1.15(10)& 5.0(10)\\
         \ch{Eu2Ir2O7/Dy2Ir2O7}&  37(5)/13(5)&  0.66(10)& 0.9(10)\\
         \ch{Y2Ir2O7}&  32(5)&  1.14(10)& 4.1(10)\\
         \ch{Y2Ir2O7/Dy2Ir2O7}&  34(5)/16(5)&  0.75(10)& 0.4(10)\\
    \end{tabular}
    \caption{XRR fitting and AFM statistic result of single and bilayer systems. AFM results were taken over a 1.5 x 1.5 $\mu$m surface}
    \label{tab:my_label}
\end{table}

\clearpage

\begin{figure}[ht]%
\centering
\includegraphics[width=0.9\textwidth]{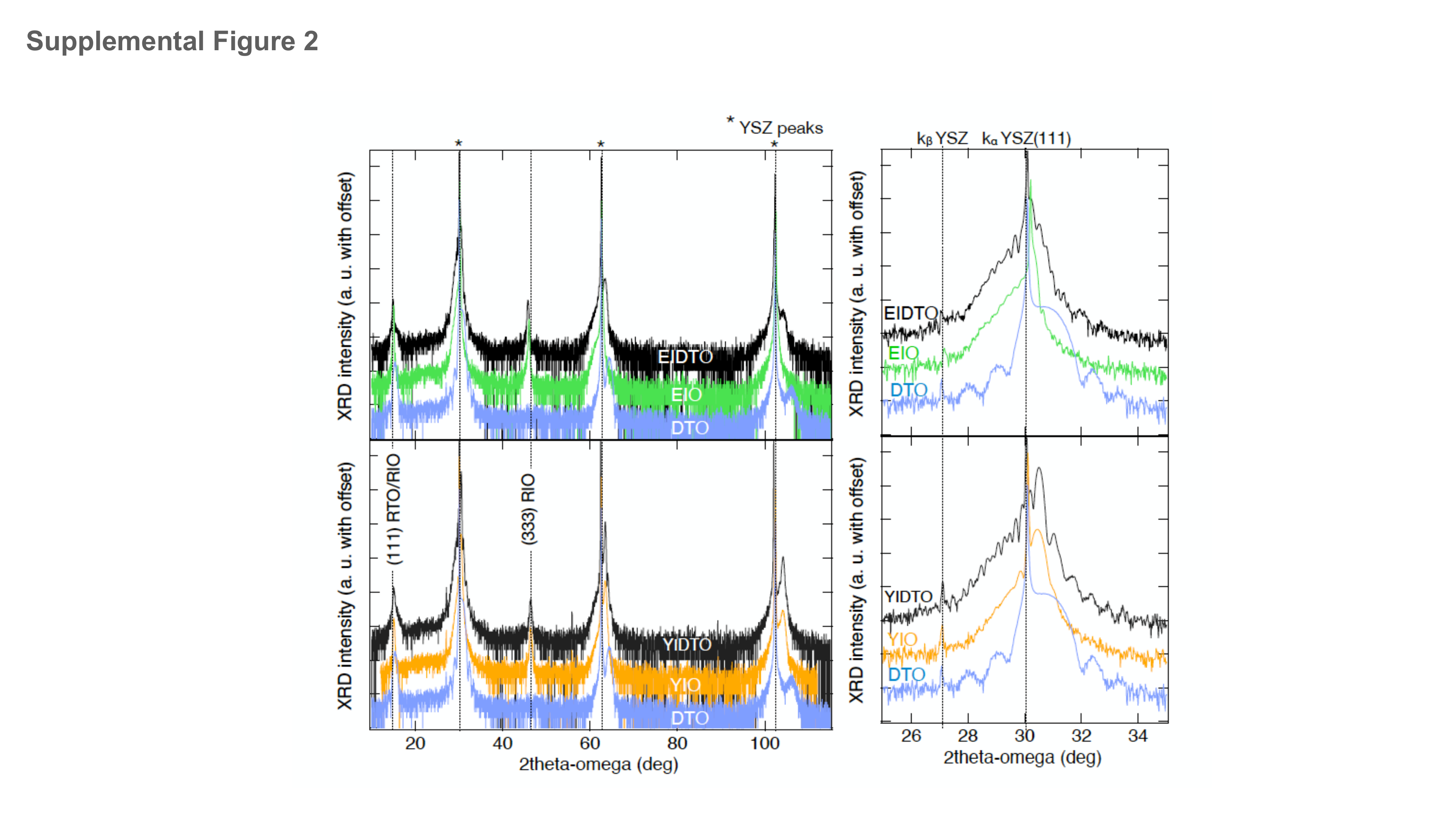}
\caption{Full-range XRD (left) and zoomed-in XRD (right) of bilayer pyrochlore systems, colored correspondingly to the previous supplemental figure 1, in comparison to the single-layer pyrochlore iridates and single layer \ch{Dy2Ti2O7}.}\label{S2}
\end{figure}

\begin{figure}[ht]%
\centering
\includegraphics[width=0.9\textwidth]{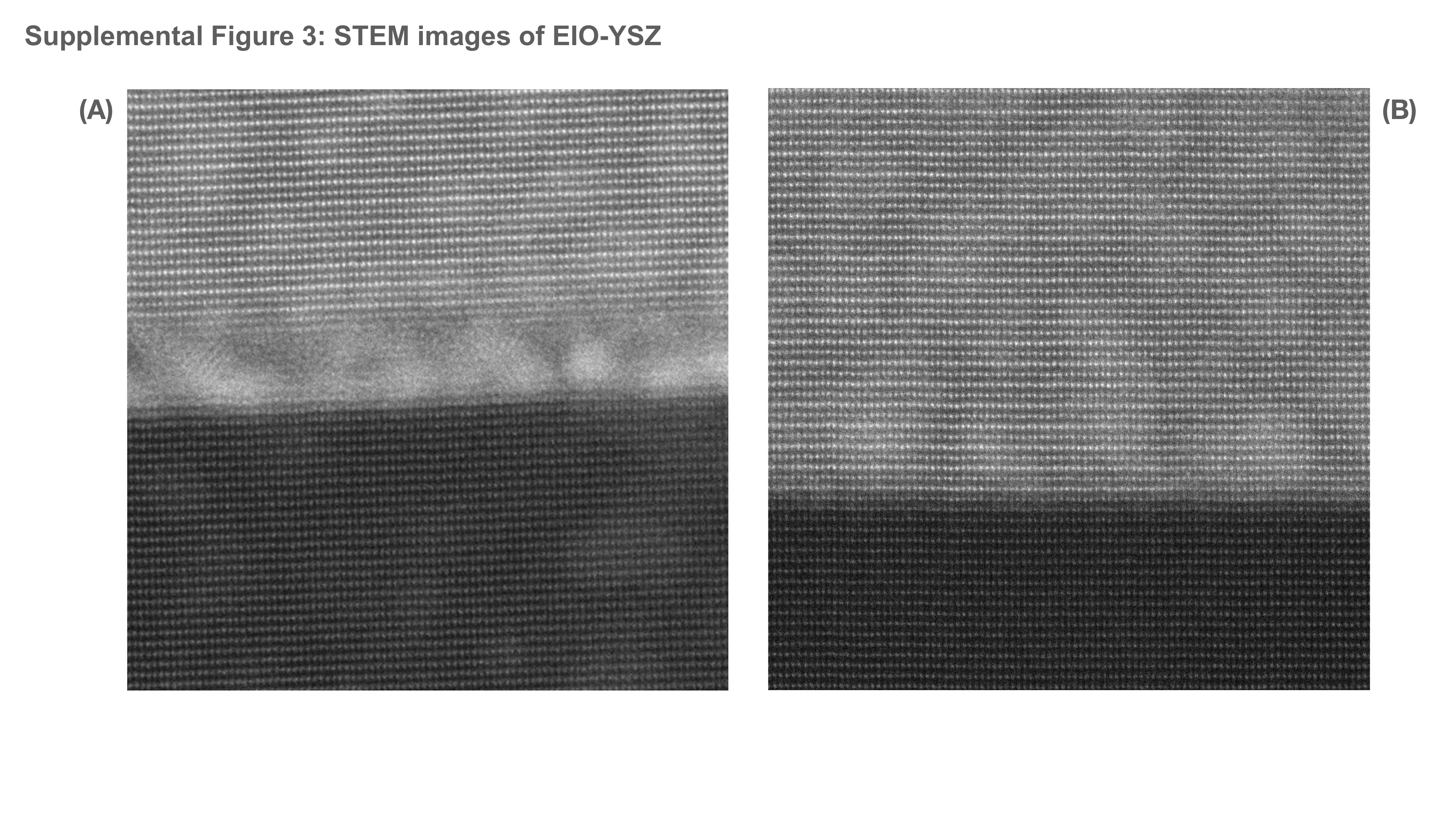}
\caption{STEM images of \ch{Eu2Ir2O7}/YSZ interfaces: (a) SPE stage ex-situ (in furnace); (b) SPE stage in-situ. }\label{S3}
\end{figure}

\begin{figure}[ht]%
\centering
\includegraphics[width=0.9\textwidth]{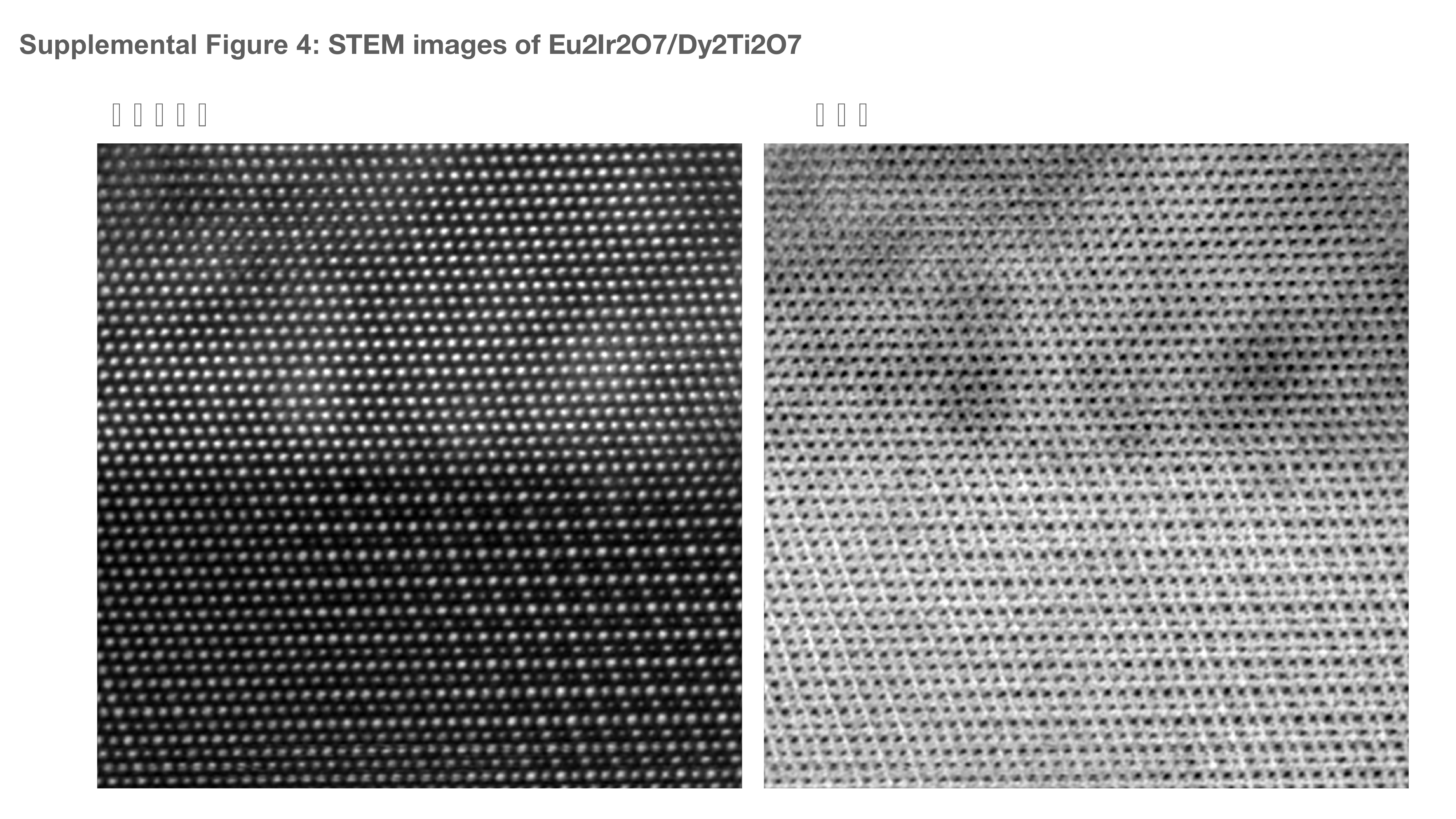}
\caption{HAADF and ABF STEM images of \ch{Eu2Ir2O7}/\ch{Dy2Ti2O7} interface.}\label{S4}
\end{figure}

\begin{figure}[ht]%
\centering
\includegraphics[width=0.9\textwidth]{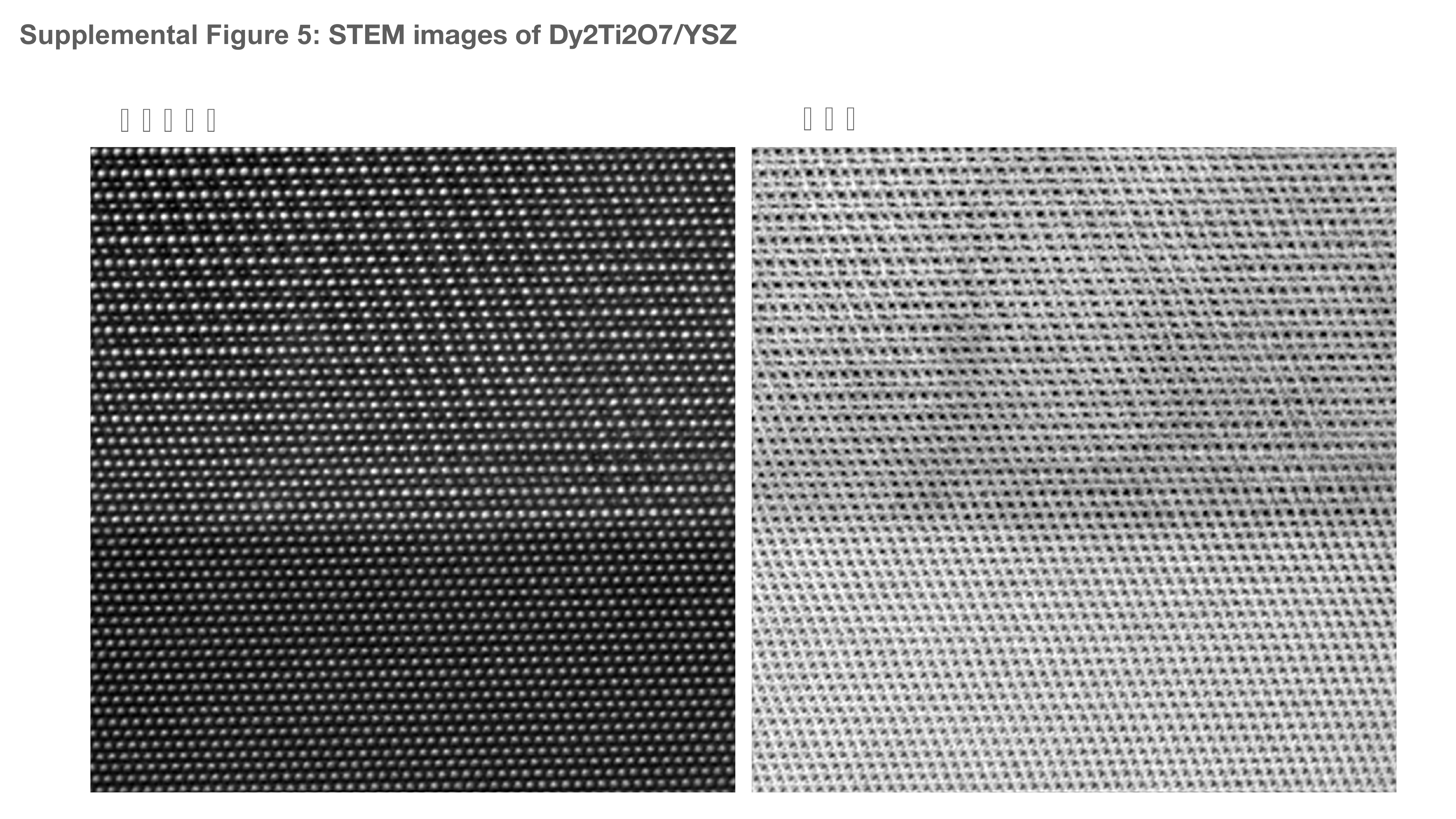}
\caption{HAADF and ABF STEM images of \ch{Dy2Ti2O7}/YSZ interface.}\label{S5}
\end{figure}

\begin{figure}[ht]%
\centering
\includegraphics[width=0.9\textwidth]{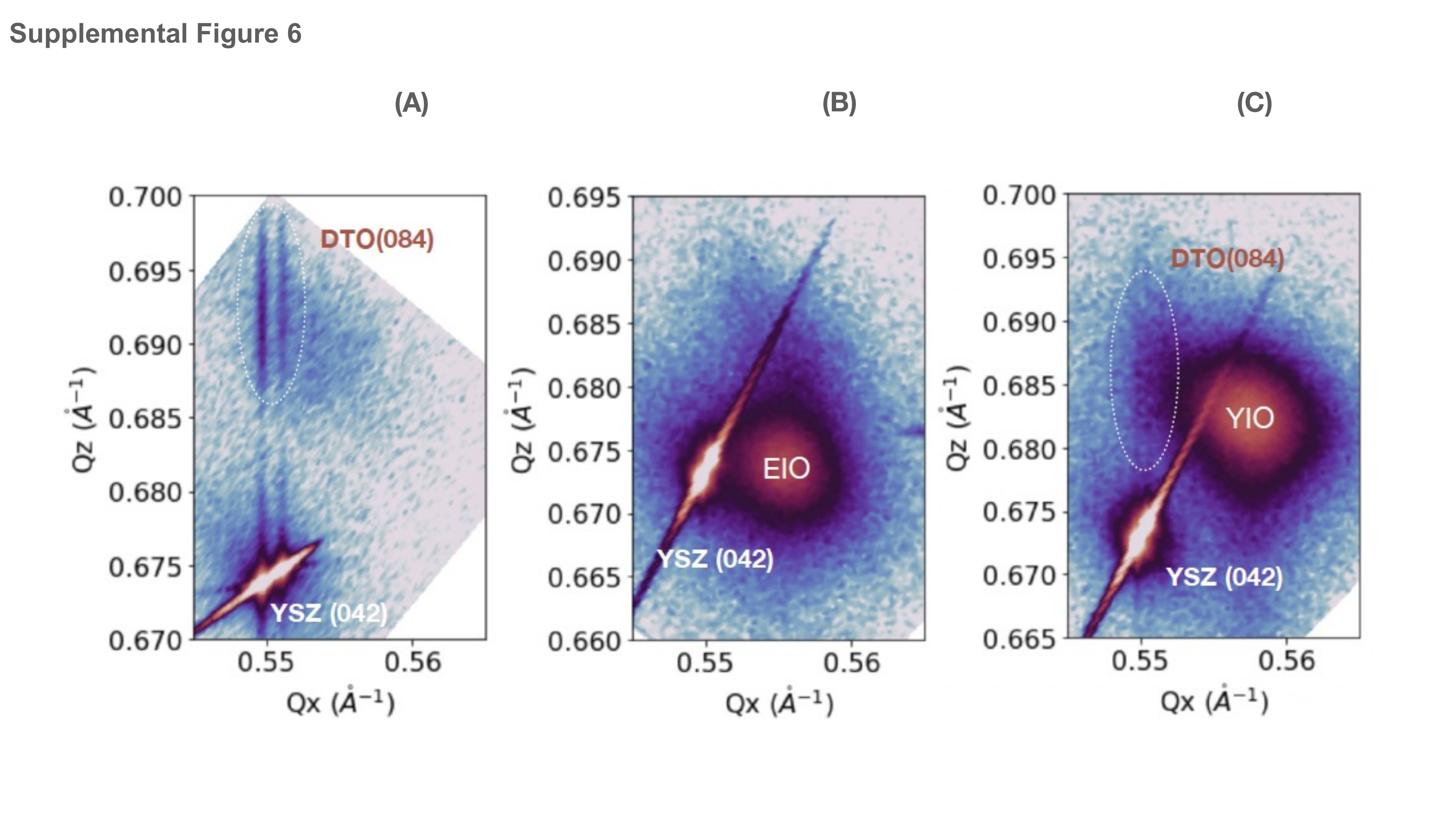}
\caption{(a) RSM of \ch{Dy2Ti2O7} single layer with thickness about 12.7 nm. (b) RSM of the \ch{Eu2Ir2O7}/\ch{Dy2Ti2O7} bilayer. (c) RSM of the \ch{Y2Ir2O7}/\ch{Dy2Ti2O7} bilayer.}\label{S6}
\end{figure}







\end{document}